\documentclass[a4paper,11pt]{article}
\usepackage{pos}
\usepackage{adjustbox}
\usepackage{multirow}
\usepackage{graphicx}
\usepackage{natbib}
\title{Gamma-ray and Optical Observations of Repeating Fast Radio Bursts with VERITAS}
 \ShortTitle{VERITAS FRB}

\author*[a]{Matthew Lundy} 

\forColl{VERITAS} 

\affiliation[a]{Department of Physics, McGill University,\\
  Montreal, QC H3A 2T8, Canada}

\emailAdd{matthew.lundy@mail.mcgill.ca}

\abstract{Fast radio burst (FRBs) are an exciting class of bright, extragalactic, millisecond radio transients. The recent development of large field-of-view (FOV) radio telescopes has caused a rapid rise in the number of identified single burst and repeating FRBs. This has allowed for the extensive multi-wavelength follow-up to search for the potential counterparts predicted by theoretical models. New observations of similar radio transients in Galactic magnetars like SGR 1935+2154 have continued to motivate the search for rapid optical and very-high-energy (VHE, >100 GeV) counterparts. Since 2016 VERITAS has engaged in an FRB observing campaign to search for the prompt optical, and VHE emission from multiple repeating FRBs. We present these new results from VERITAS observations of five repeating sources including data taken simultaneously with bursts observed by the CHIME radio telescope.}

\FullConference{37$^{\rm{th}}$ International Cosmic Ray Conference (ICRC 2021)\\
		July 12th -- 23rd, 2021\\
		Online -- Berlin, Germany}

\begin{document}
\maketitle

\section{Introduction}
 Fast Radio Bursts (FRBs) are bright ($50$ mJy - 100 Jy \cite{2019A&ARv..27....4P}), $\sim$ms duration radio bursts, occurring in extra-galactic environments \cite{2017ApJ...834L...7T}, from an unknown class of progenitor \cite{2021SCPMA..6449501X}. For a recent review of the fundamentals of FRBs see Petroff, 2019 \cite{2019A&ARv..27....4P}. Although the estimated all-sky rate of FRBs is $\sim10^3$ per day \cite{2018MNRAS.475.1427B}, the rapid nature of these transients and the small field-of-view (FOV) of traditional radio telescopes meant that only $\sim80$ bursts were discovered within the first decade of study \cite{2019ARA&A..57..417C}.  The arrival of large FOV radio survey telescopes have caused a massive rise in the measured number of both non-repeating and repeating FRBs \cite{2019ApJ...885L..24C}. In the most recent catalog from the Canadian Hydrogen Intensity Mapping Experiment (CHIME) 535 bursts were published, increasing the known population by a factor of four \cite{2021arXiv210604352T}. Many notable FRBs have also emerged in the past year due to CHIME and other efforts, including an FRB localized to a globular cluster in M81 \cite{2021arXiv210511445K}, a periodic repeating FRB \cite{2020Natur.582..351C}, and thirteen well localized FRBs (such as \cite{2017ApJ...834L...7T,2017Natur.541...58C,2019Sci...365..565B}). The rise of CHIME in the FRB field has also allowed for many novel multiwavelength opportunities to emerge that were previously unfeasible due to the rarity and sporadic nature of these transients. 

Multi-wavelength searches are critical for validating the predictions of many theoretical progenitor models. To date there has been no unambiguous multi-wavelength counterpart observed for an extra-galactic FRB. An exciting recent observation of FRB-like pulses from SGR 1935+2154 has provided some of the first evidence supporting the class of magnetar models for FRBs \cite{2020Natur.587...54C,2020Natur.587...59B,2020ApJ...898L..29M}. Although these short FRB-like pulses, measured simultaneously with X-ray pulses from the source, were 3 orders of magnitude more luminous than typical Soft Gamma-ray Repeater (SGR) pulses, there still remains an ``energy gap'' of approximately 2 orders of magnitude between SGRs and the weakest FRBs \cite{2020ApJ...899L..27M}. It is still unclear, in the magnetar model, what accounts for the large discrepancy and the multiwavelength implications of this difference. 

Due to the high bolometric luminosity of these sources, gamma-ray emission has been expected by some models and VHE neutrino emission has also been theorized \cite{2014MNRAS.442L...9L,2013arXiv1307.4924P}. Imaging Atmospheric Cherenkov Telescopes (IACTs) like VERITAS are uniquely poised to take advantage of this. In addition to the high-energy capabilities, the rapid photosensors and large mirror area also make IACTs an ideal instrument to investigate the rapid optical emission from these sources \cite{2019NatAs...3..511B}. In these proceedings we report on the status of a parallel optical/VHE program at VERITAS investigating FRBs. We will present results of VERITAS observations of 5 repeaters in the 2020 season and discuss the status of the 2021 observations as well as VERITAS future plans.

\section{Instrumentation}

The CHIME radio telescope is located at the Dominion Radio Observatory near Penticton, British Columbia, Canada \cite{2018ApJ...863...48C}. The telescope is composed of four 20 m by 100 m dishes aligned in the the N/S direction. A linear array of 256 dual-polarization antennas monitor a FOV of 120$^\circ$ in the North/South direction and $1.3-2.5^\circ$ in the East/West (due to the significant frequency and declination dependence) \cite{2017arXiv170204728N}. This combined with CHIME's geographic position means that the FRB detection instrument will monitor the entire Northern sky every day (above a declination of $11^\circ$). CHIME monitors a frequency range from 400 MHz to 800 MHz and the beam width FWHM varies as a function of the frequency from $40'-20'$ \cite{2018ApJ...863...48C}. The localization of bursts is of a comparable magnitude but the exact structure is complicated due to the presence of factors like side lobe contamination.

The VERITAS observatory consists of four 12-meter diameter telescopes located at the Fred Lawrence Whipple Observatory, in Amado, Arizona, USA \cite{2006APh....25..391H,2015ICRC...34..771P}.  VERITAS is designed to observe gamma-rays in the very-high-energy (VHE) regime, between 85 GeV and 30 TeV. Gamma-rays are measured through observations of Cherenkov light produced in the particle cascades of gamma-ray photons in the atmosphere. The gamma-ray sensitivity of VERITAS allows for the detection of a source at 10\% of the Crab flux in $\sim$25 minutes.

Traditionally only the triggered photomultiplier tube (PMT) pulses are saved for shower reconstruction but upgrades in 2016 and 2018 have allowed for VERITAS to monitor the DC light level in a subset of pixels. This allows for optical photometry to be done with a portion of the VERITAS camera while normal gamma-ray observations are underway. The VERITAS camera is composed of 499 photomultiplier tubes with a $0.15^\circ$ pixel field of view, matching the size of the optical point spread function of the telescope \cite{2006APh....25..391H}. Attached to these is a commercial DATAQ DI-710-ELS DC voltage data logger with the ability to sample the PMT signal at a rate of 4,800 Hz at 14 bit resolution. Changing configurations can improve the timing resolution of the telescope while sacrificing the field of view being monitored as the sample rate is divided among the monitored pixels. Recent FRB observations use an updated configuration where two pixels, a central pixel and a background pixel, are monitored on three telescopes. The final telescope monitors four pixels that span a wider field of view, selected to isolate background signals. This telescope samples at a cadence of 1.2 kHz, whereas the other three telescopes sample at a higher rate of 2.4 kHz. The VERITAS PMTs are sensitive from approximately 250 nm to 550 nm with the peak sensitivity at $\sim350$ nm.

The relative geographic location of CHIME and VERITAS allows for a unique monitoring campaign. Since the sites of the two telescopes are only separated by 8.7$^\circ$ in longitude, if VERITAS observes sources $\sim30$ minutes past culmination, they will be transiting through the CHIME field of view. The duration of an object's transit through the CHIME FOV is dependent on the declination of the source and can range from $\sim5-30$ minutes. When VERITAS takes targeted observations of repeaters it guarantees simultaneous radio data during the gamma-ray observations. In contrast to other high energy follow-up efforts of FRBs, that carry a delay, the gamma-ray and optical data will overlap with the arrival time of the FRB detection and also have several minutes of data surrounding the event.

\section{Repeater Observations}

FRB repeaters have been selected for monitoring based on factors that maximize the potential VHE and optical emission. The repeaters observed up until summer 2020 are presented here with a summary of the observations presented in Table \ref{tab:veritabl}. Observations prior to 2019 were taken in 30-minute exposures but beginning in 2019 observations were modified such that VERITAS only observed the source while it was transiting the CHIME FOV. These observations were taken in "ON" mode, where the telescope FOV is centered on the source. 

For the 2019 season, FRB repeaters were added to the campaign as they were discovered, without any discrimination. In later seasons repeaters were selected to optimize the likelihood of observing a multiwavelength counterpart. We sought to select a mixture of repeaters with low dispersion measure (the integrated electron density which acts as a proxy for distance), high burst rates in CHIME, low VERITAS zenith angles at culmination (which lowers the VERITAS gamma-ray energy threshold), high CHIME signal-to-noise ratio (a rough proxy for fluence), and precise localization. Notable repeaters like FRB 121102 were dropped in later seasons based on this evaluation. 

In addition to the previously reported limits from bursts from FRB 121102 \cite{2019ICRC...36..698H}, VERITAS has also collected data during three contemporaneous or near-contemporaneous bursts with CHIME from the repeater FRB 180916.J0158+65. The properties of these bursts are shown in Table \ref{tab:Bursttimes}.

\section{Analysis and Preliminary Results}

\begin{figure*}[ht]
  \begin{adjustbox}{addcode={\begin{minipage}{\width}}{\caption{%
      VERITAS gamma-ray significance maps centered on the best fit position of the FRBs. The FRBs shown are; a) FRB 121102, b) FRB 180814.J0422+73, c) FRB 180916.J0158+65, d) FRB 181030.J1054+73, e) FRB 190116.J1249+27. The color indicates the significance of a gamma-ray source located at the sky position. FRBs that have sub-arcminute localization have the FRB position represented by a black cross. Other FRBs have the CHIME 95\% contour interval shown. For FRB 190116.J1249+27 the data is centered on the western lobe of the 95\% confidence region. 
      \label{fig:sigmaps}}\end{minipage}},rotate=90,center}
      \includegraphics[width=1.2\textwidth]{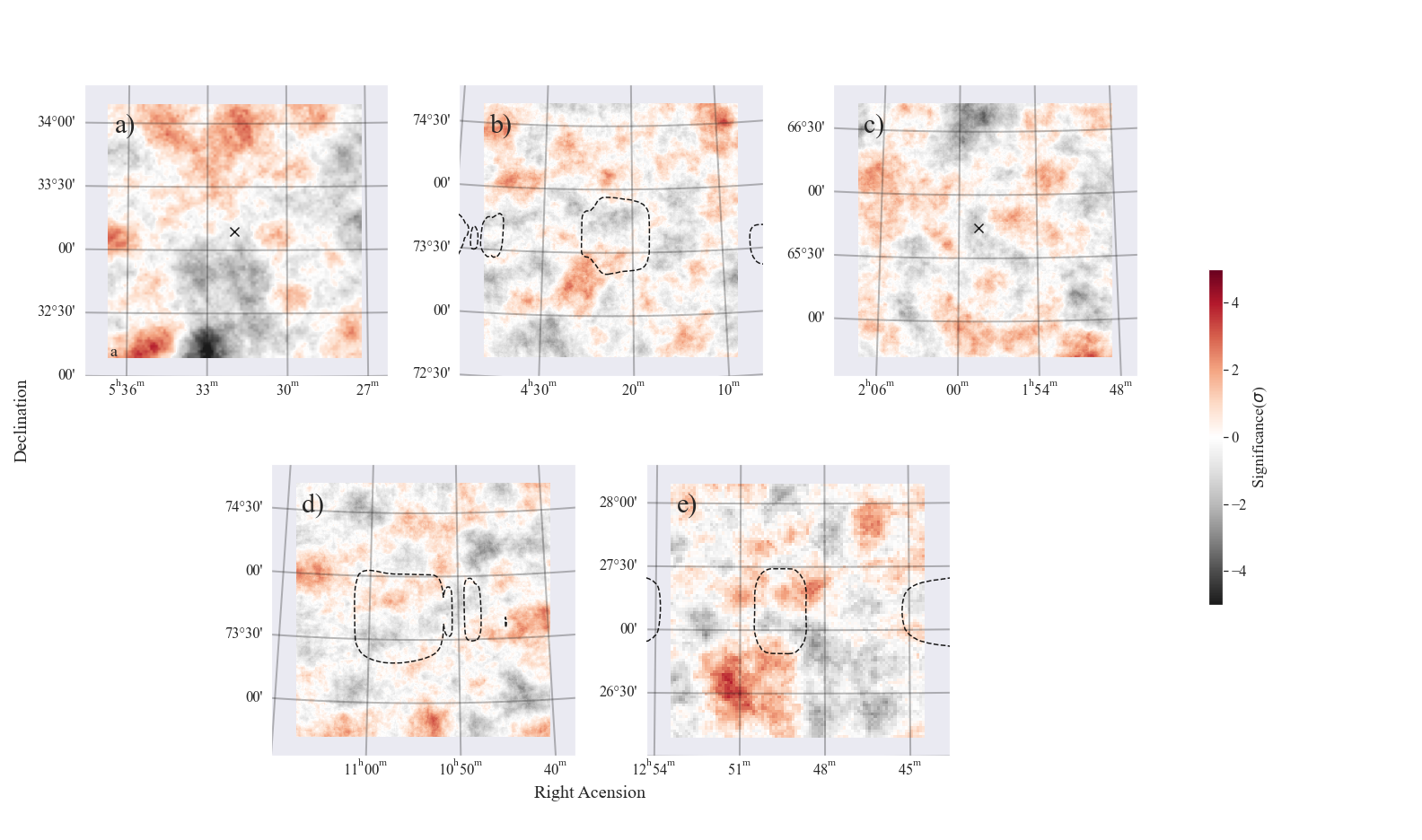}%
  \end{adjustbox}
  
\end{figure*}

An analysis of the gamma-ray data was performed using the Ring Background Method (RBM), where observations were taken in "ON" mode and backgrounds are estimated using a events taken from a ring around the source location \cite{2008ICRC....3.1385C}. A series of gamma-hadron cuts that optimized for \textit{soft} sources (a spectral index -3.5 to -4.0) were applied to the data. For all repeaters, the first analysis that was performed was an integrated search for emission around the source location using the cumulative exposure. The events were smoothed using the gamma-ray point spread function of VERITAS and the results are shown in Figure \ref{fig:sigmaps}. No significant excess was found at the most likely FRB position, the values of which are shown in Table \ref{tab:veritabl}, or within the larger error region of any of the repeaters. 

\begin{figure*}
    \centering
    \includegraphics[width=\textwidth]{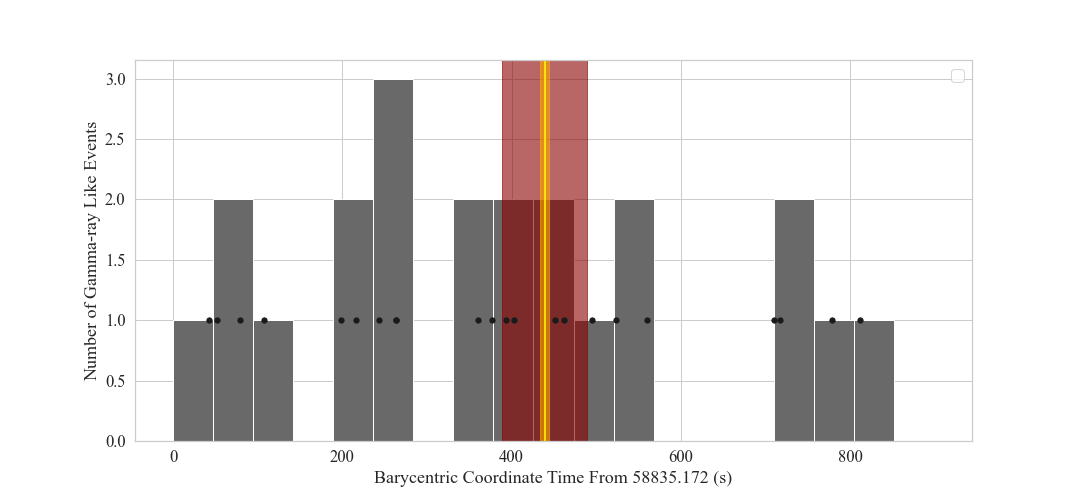}
    \caption{Burst analysis time series for the simultaneous FRB. A gray histogram of the number of gamma-ray counts is shown. Burst in Black points show the arrival time of the gamma-like events. The 10 and 100 s windows are shown in yellow and red respectively.}
    \label{fig:lclc}
\end{figure*}

For runs that lie within 1 hour of a known repeater burst, the events in the ON region are extracted from the runs and an average background rate is extracted from the RBM analysis. Then, logarithmically spaced windows of 0.1 s, 10 s, 100 s, and 1000 s are assessed to search for evidence of significant emission during the run coincident with the arrival time of the barycentric, dispersion-corrected burst. The arrival times of the near-by events for FRB J180916.J0158+65 are shown in Table \ref{tab:Bursttimes}. No significant excess is seen in any of the windows surrounding these bursts, one of which is shown in Figure \ref{fig:lclc}.

If a repeater has been localized by a smaller field of view telescope then an optical analysis can be performed at the location of the source. Prior to localization, the large optical pixel size of VERITAS means that source confusion and large non-astrophysical backgrounds leaves the nature of most transients ambiguous.

{
\centering
\begin{table*}
    \caption{VERITAS Observation Log for FRB repeaters. The significances are calculated using Equation 17 in \cite{1983ApJ...272..317L} at the best fit position of the FRB.} 
    
    \begin{tabular}{lllll}
    \hline
    \hline

    FRB Name & Exposure (min) & On Counts & Off Counts & Significance($\sigma$) \\ \hline
    FRB 121102 & 1216.64 & 1681  & 14134  & -0.61\\
    FRB 180814.J0422+73 & 1013.22 & 966 & 8955 & -0.62\\
    FRB 180916.J0158+65 & 397.45 & 522 & 4907 & -0.06 \\
    FRB 181030.J1054+73	& 226.26 & 277  & 2650  & -0.33 \\
    FRB 190116.J1249+27	&  45.00  & 111 & 768 &  \textrm{ }0.83 \\
    \hline
    \hline
    \end{tabular}
    
\label{tab:veritabl}

\end{table*}
}

\noindent
{
\centering
\begin{table*}
    \caption{Fast radio bursts from FRB J180916.J0158+65 with CHIME and VERITAS overlap. All CHIME information taken from \url{//www.chime-frb.ca/repeaters}}
    \centering
    \begin{tabular}{p{5.0cm}p{4.5cm}p{3.5cm}}
    \hline
    \hline

     Time of CHIME FRB (UTC) & VERITAS Start of Observation (UTC)& Duration of VERITAS Observation (s) \\ \hline
     2019-12-18 04:09:27.633  & 2019-12-18 04:01:59.29 &900\\
     2019-10-30 07:33:56.995676 & 2019-10-30 07:15:37.69 &811\\
     2020-01-20 01:49:14.068 & 2020-01-20 02:03:01.38&1800 \\
    \hline
    \hline
    \end{tabular}
    
\label{tab:Bursttimes}

\end{table*}
}

\section{Discussion and Future Plans}

\subsection{2020/2021 Season}
During the 2020/2021 season we updated our repeater target list based on new available repeaters from CHIME. A total of 49.3 hours of data was collected as a part of this campaign. Data were collected on previous repeaters as well as the addition of FRB 190213.J02+20, FRB 181119.J12+65, FRB 190303.J1353+48, FRB 190212.J18+81, and M 81. Coordinates were taken from the CHIME repeater catalog at the most probable location if no precise localization had been obtained by a small FOV radio telescope. This data will be combined with the preliminary results presented here in a future publication. Additional bursts will also be presented.

\subsection{VOEvents}

VERITAS begun monitoring VOEvents broadcast by CHIME \cite{2017arXiv171008155P} towards the end of March 2021. CHIME releases a realtime stream of FRB candidates in VOEvent format to subscribers of their Comet event broker. VOEvents offer the opportunity for realtime follow-up of not precisely localized repeaters. To date the rate of FRB-like events has made realtime follow-up of events impractical. The system is currently integrated as a passive monitor with a local event parser comparing the live VERITAS pointing to the location of events and sending an alert if there is been a burst whose error region coincides with the FOV of VERITAS.  A future plan may include a short-duration campaign of follow-up, but a constant integration of events into the GRB pipeline of VERITAS. An assessment of this possibility in underway.

\subsection{Conclusion}

We have presented the current status of the VERITAS fast radio burst campaign. As we learn more about FRBs and as the CHIME instrument matures, the ability for more detailed multi-wavelength searches continues to grow. As more sources continue to be localized the possibility of coincident optical observations with VERITAS becomes more likely. The release of the CHIME single burst FRB catalog has also allowed investigations of the population of non-repeating FRBs using archival VERITAS data.

\section{Acknowledgments} 

\noindent For VERITAS acknowledgments see: \url{https://veritas.sao.arizona.edu/}

\noindent Additionally this research was undertaken thanks in part to funding from the Canada First Research Excellence Fund through the Arthur B. McDonald Canadian Astroparticle Physics Research Institute. Thanks should also be given to the whole CHIME FRB team at the McGill Space Institute.  
\medskip

\small
\bibliography{bib}

\clearpage
\section*{Full Authors List: \Coll\ Collaboration}

\scriptsize
\noindent
C.~B.~Adams$^{1}$,
A.~Archer$^{2}$,
W.~Benbow$^{3}$,
A.~Brill$^{1}$,
J.~H.~Buckley$^{4}$,
M.~Capasso$^{5}$,
J.~L.~Christiansen$^{6}$,
A.~J.~Chromey$^{7}$, 
M.~Errando$^{4}$,
A.~Falcone$^{8}$,
K.~A.~Farrell$^{9}$,
Q.~Feng$^{5}$,
G.~M.~Foote$^{10}$,
L.~Fortson$^{11}$,
A.~Furniss$^{12}$,
A.~Gent$^{13}$,
G.~H.~Gillanders$^{14}$,
C.~Giuri$^{15}$,
O.~Gueta$^{15}$,
D.~Hanna$^{16}$,
O.~Hervet$^{17}$,
J.~Holder$^{10}$,
B.~Hona$^{18}$,
T.~B.~Humensky$^{1}$,
W.~Jin$^{19}$,
P.~Kaaret$^{20}$,
M.~Kertzman$^{2}$,
D.~Kieda$^{18}$,
T.~K.~Kleiner$^{15}$,
S.~Kumar$^{16}$,
M.~J.~Lang$^{14}$,
M.~Lundy$^{16}$,
G.~Maier$^{15}$,
C.~E~McGrath$^{9}$,
P.~Moriarty$^{14}$,
R.~Mukherjee$^{5}$,
D.~Nieto$^{21}$,
M.~Nievas-Rosillo$^{15}$,
S.~O'Brien$^{16}$,
R.~A.~Ong$^{22}$,
A.~N.~Otte$^{13}$,
S.~R. Patel$^{15}$,
S.~Patel$^{20}$,
K.~Pfrang$^{15}$,
M.~Pohl$^{23,15}$,
R.~R.~Prado$^{15}$,
E.~Pueschel$^{15}$,
J.~Quinn$^{9}$,
K.~Ragan$^{16}$,
P.~T.~Reynolds$^{24}$,
D.~Ribeiro$^{1}$,
E.~Roache$^{3}$,
J.~L.~Ryan$^{22}$,
I.~Sadeh$^{15}$,
M.~Santander$^{19}$,
G.~H.~Sembroski$^{25}$,
R.~Shang$^{22}$,
D.~Tak$^{15}$,
V.~V.~Vassiliev$^{22}$,
A.~Weinstein$^{7}$,
D.~A.~Williams$^{17}$,
and 
T.~J.~Williamson$^{10}$\\
\noindent
$^{1}${Physics Department, Columbia University, New York, NY 10027, USA}
$^{2}${Department of Physics and Astronomy, DePauw University, Greencastle, IN 46135-0037, USA}
$^{3}${Center for Astrophysics $|$ Harvard \& Smithsonian, Cambridge, MA 02138, USA}
$^{4}${Department of Physics, Washington University, St. Louis, MO 63130, USA}
$^{5}${Department of Physics and Astronomy, Barnard College, Columbia University, NY 10027, USA}
$^{6}${Physics Department, California Polytechnic State University, San Luis Obispo, CA 94307, USA} 
$^{7}${Department of Physics and Astronomy, Iowa State University, Ames, IA 50011, USA}
$^{8}${Department of Astronomy and Astrophysics, 525 Davey Lab, Pennsylvania State University, University Park, PA 16802, USA}
$^{9}${School of Physics, University College Dublin, Belfield, Dublin 4, Ireland}
$^{10}${Department of Physics and Astronomy and the Bartol Research Institute, University of Delaware, Newark, DE 19716, USA}
$^{11}${School of Physics and Astronomy, University of Minnesota, Minneapolis, MN 55455, USA}
$^{12}${Department of Physics, California State University - East Bay, Hayward, CA 94542, USA}
$^{13}${School of Physics and Center for Relativistic Astrophysics, Georgia Institute of Technology, 837 State Street NW, Atlanta, GA 30332-0430}
$^{14}${School of Physics, National University of Ireland Galway, University Road, Galway, Ireland}
$^{15}${DESY, Platanenallee 6, 15738 Zeuthen, Germany}
$^{16}${Physics Department, McGill University, Montreal, QC H3A 2T8, Canada}
$^{17}${Santa Cruz Institute for Particle Physics and Department of Physics, University of California, Santa Cruz, CA 95064, USA}
$^{18}${Department of Physics and Astronomy, University of Utah, Salt Lake City, UT 84112, USA}
$^{19}${Department of Physics and Astronomy, University of Alabama, Tuscaloosa, AL 35487, USA}
$^{20}${Department of Physics and Astronomy, University of Iowa, Van Allen Hall, Iowa City, IA 52242, USA}
$^{21}${Institute of Particle and Cosmos Physics, Universidad Complutense de Madrid, 28040 Madrid, Spain}
$^{22}${Department of Physics and Astronomy, University of California, Los Angeles, CA 90095, USA}
$^{23}${Institute of Physics and Astronomy, University of Potsdam, 14476 Potsdam-Golm, Germany}
$^{24}${Department of Physical Sciences, Munster Technological University, Bishopstown, Cork, T12 P928, Ireland}
$^{25}${Department of Physics and Astronomy, Purdue University, West Lafayette, IN 47907, USA}

\end{document}